\documentclass[openacc]{rstransa}
\usepackage{slashed}

\newcommand{\be}{\begin{equation}}
\newcommand{\ee}{\end{equation}}
\newcommand{\ba}{\begin{array}}
\newcommand{\ea}{\end{array}}
\newcommand{\bea}{\begin{eqnarray}}
\newcommand{\eea}{\end{eqnarray}}
\newcommand{\sss}{\scriptscriptstyle}
\newcommand{\R}{{\sss R}}
\renewcommand{\L}{{\sss L}}
\def\sfrac#1#2{{\textstyle{#1\over #2}}}

\begin{document}

\title{Is electroweak baryogenesis dead?}

\author{
James M.\ Cline$^{1,2}$}

\address{$^{1}$CERN, Theoretical Physics Department, Geneva,
Switzerland\\
$^{2}$Department of Physics, McGill University,
3600 Rue University, Montr\'eal, Qu\'ebec, Canada H3A 2T8 }

\subject{particle physics, cosmology}

\keywords{electroweak baryogenesis, electroweak phase transition, dark matter}

\corres{James M.\ Cline\\
\email{jcline@physics.mcgill.ca}}

\begin{abstract}
Electroweak baryogenesis is severely challenged in its traditional settings: the Minimal Supersymmetric Standard
Model, and in more general two Higgs doublet models.  Fine tuning of parameters is required, or large couplings leading to
a Landau pole at scales just above the new physics introduced.  The situation is somewhat better in models with a singlet
scalar coupling to the Higgs so as to give a strongly first order phase transition due to a tree-level barrier, but 
even in this case no UV complete models had been demonstrated to give successful baryogenesis.  Here we point out some 
directions that overcome this limitation, by introducing a new source of CP violation in the couplings of the singlet
field.  A model of electroweak baryogenesis requiring no fine tuning and consistent to scales far above 1 TeV is
demonstrated, in which dark matter plays the leading role in creating a CP asymmetry that is the source of the baryon
asymmetry. 
\end{abstract}


\begin{fmtext}
\section{Introduction}\label{intro}

Readers old enough to remember Hinchliffe's rule \cite{Peon:1988kx} will guess that the answer to the question of the
title given here is ``no.''   But before elaborating the challenges faced by electroweak baryogenesis, it is well to remind the reader why they should
care.   The preferred paradigm of many physicists for creating the baryon asymmetry of the universe (BAU) is leptogenesis,
since it is a feasible mechanism that comes almost for free just by invoking the seesaw mechanism for neutrino masses.
However with the notable exception of resonant leptogenesis using low-scale right-handed neutrinos \cite{Pilaftsis:2003gt}, this appealing
idea may never be experimentally verifiable, since it relies upon new physics at the scale of the heavy neutrinos,
$\sim 10^{10}\,$GeV.  

\end{fmtext}
\maketitle

Electroweak baryogenesis (EWBG) is by design highly testable at colliders since it relies upon new physics at the scale of the
electroweak phase transition.  In principle, we expect that it should be verified during the LHC era. One might question whether
that test has essentially been already carried out now, with a negative conclusion, hence the title of this contribution.  Here
we will present one class of examples to the contrary, but in fact there are also others that have been discussed at this
meeting \cite{Baldes:2016gaf}.  The model we focus upon is based upon a two-step phase transition in which a singlet field first
condenses, before making a strongly first order transition to the electroweak symmetry breaking (EWSB) vacuum.  It is also
possible to get a strong EWSB transition using a second field that transforms nontrivially under the standard model SU(2)$_L$
gauge symmetry \cite{Patel:2012pi,Blinov:2015sna,Inoue:2015pza}.

\begin{figure}[t]
\centering
\centerline{\includegraphics[width=2.5in]{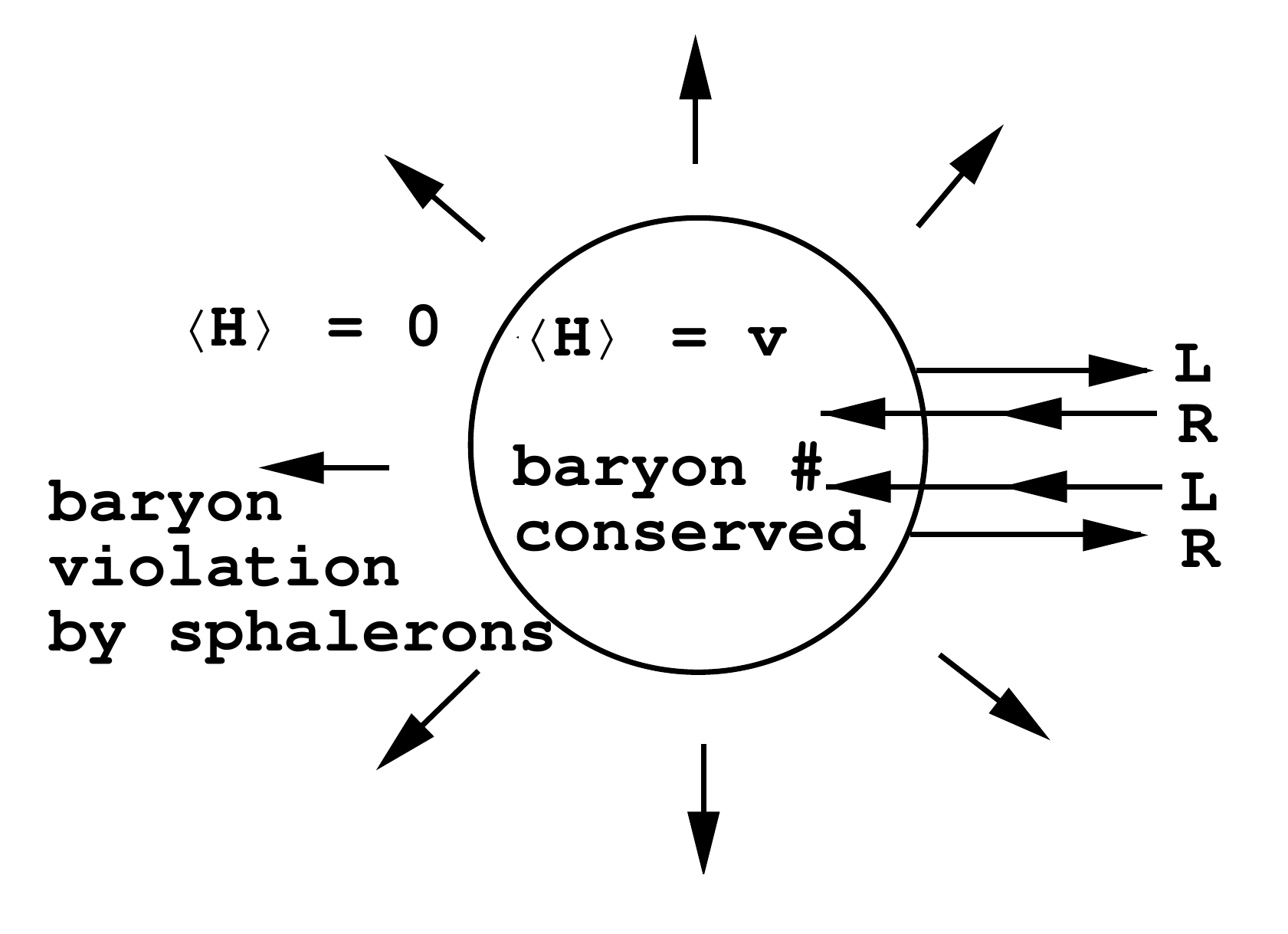}\hfil\raisebox{0ex}{\includegraphics[width=3.5in]{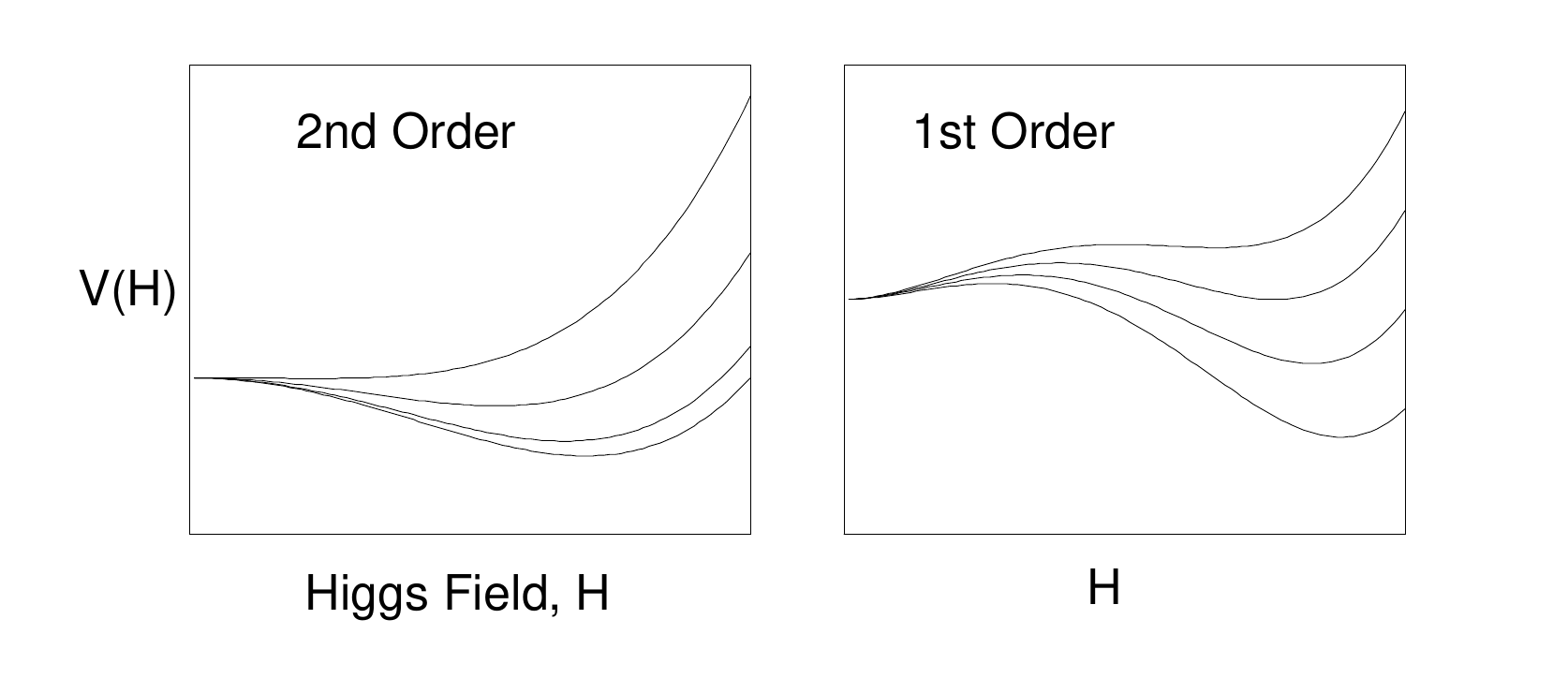}}}
\centerline{$\qquad$ (a) $\qquad\qquad\qquad\qquad\qquad\qquad\qquad\qquad\qquad\qquad\qquad$ (b) $\qquad\qquad\qquad$}
\caption{(a) The essential ingredients of electroweak baryogenesis, surrounding a bubble nucleated
during a first order electroweak phase transition.  (b) Evolution of the Higgs potential with temperature for
a second or first order phase transition.}
\label{bubble}
\end{figure}

For completeness we briefly recapitulate the essential ingredients of electroweak baryogenesis 
\cite{Kuzmin:1985mm,Cohen:1990it,Turok:1990in}, summarized in fig.\ \ref{bubble}(a).  If the electroweak phase transition
(EWPT) is first order, bubbles of the broken phase with nonvanishing Higgs VEV $v$ will nucleate and grow.  Standard model fermions
should interact with the bubble walls in a CP-violating manner so as to produce a chiral asymmetry---an excess of left-handed
versus right-handed fermions in front of the wall.  In this region, baryon-violating sphaleron interactions are in 
thermal equilibrium, and try to erase the chiral asymmetry, converting it into a baryon asymmetry.  These baryons eventually
fall inside the expanding bubble, and are safe from washout by sphalerons inside the bubble as long as 
\be
v > 1.1\, T,
\label{vot}
\ee
the condition for the sphaleron interactions to go out of equilibrium \cite{Moore:1998swa}.  To achieve a first order phase
transition, the Higgs potential must develop a barrier between the symmetric and symmetry-breaking minima,
as illustrated in the rightmost of fig.\ \ref{bubble}(b).

There are two main difficulties for getting successful EWBG.  The first is that condition (\ref{vot}) is hard to 
achieve from a barrier generated by thermal corrections to the effective potential.  The most important such correction,
in an expansion of field-dependent masses over temperature, is the cubic term
\be
	\Delta V(h) \ni -{T\over 12\pi} \sum_i (m_i^2(h))^{3/2}= 
-{T\over 12\pi} \sum_i (m_{i,0}^2 + g_i^2 h^2 + c_i T^2)^{3/2}
\ee
In the absence of the bare mass and thermal correction, this would have a pure cubic form,
$(g_i h)^3 T/12\pi$, leading to the desired barrier in the potential.  But if $m_{i,0}^2 + c_i T^2$ is not small
or the coupling $g_i$ is weak, then the barrier is low and leads to a smaller VEV than required by (\ref{vot}).
To overcome this one typically needs to choose large couplings and tune the bare mass.

The second difficulty is in getting strong enough CP-violation in the interactions of fermions with the bubble wall.
The new CP-violating interactions are often highly constrained by experimental limits on electric dipole moments of
the neutron, electron, and certain atomic nuclei.

Beyond these intrinsically physical challenges, it is also technically difficult to accurately predict the baryon asymmetry 
for a given model.  One should determine the actual nucleation temperature $T_n$ when the first order
transition occurs, rather than just the critical temperature $T_c$ when the true and false vacua of the Higgs potential become
degenerate (and the potential has the simple form $V = \lambda h^2 (h-v_c)^2$).  Then at $T_n$ the profile of the bubble wall 
$h(z)$ should be found by solving the Higgs field equation
of motion (and in a two-field model, for example with a singlet $s$, one must solve for $s(z)$ as well).  Many treatments
use an approximation for the wall profile by an idealized form $h(z) = \sfrac12(1-\tanh(z/L))$ with
and $L = \sqrt{2/\lambda v_c^2}$ that is correct at $T_c$, but an accurate determination may require better
knowledge of the shape and thickness of the wall.  

Moreover the BAU depends upon how fast the wall is moving, $v_w$, which is hard to compute, depending upon the friction exerted
by the particles in the plasma on the wall \cite{Moore:1995ua}, without which it would accelerate to the speed of light, known
as a runaway wall.  For low speeds, the BAU is relatively insensitive to $v_w$, but it is understood that as $v_w$ approaches
the sound speed of the plasma $\cong 1/\sqrt{3}$, baryon production  will be suppressed because nothing can diffuse in front of
the wall (however a quantitative study of this effect is lacking in the literature).  Fast walls are correlated with
very strong phase transitions \cite{Kozaczuk:2015owa,Kurup:2017dzf}, so although the latter is desired to  preserve the baryon
asymmetry, eq.\ (\ref{vot}), there is the risk of not making enough in the first place.  The criteria for producing runaway
walls are discussed in refs.\ \cite{Bodeker:2009qy,Bodeker:2017cim}.

\begin{figure}[t]
\centering
\centerline{\raisebox{5ex}{\includegraphics[width=1.5in]{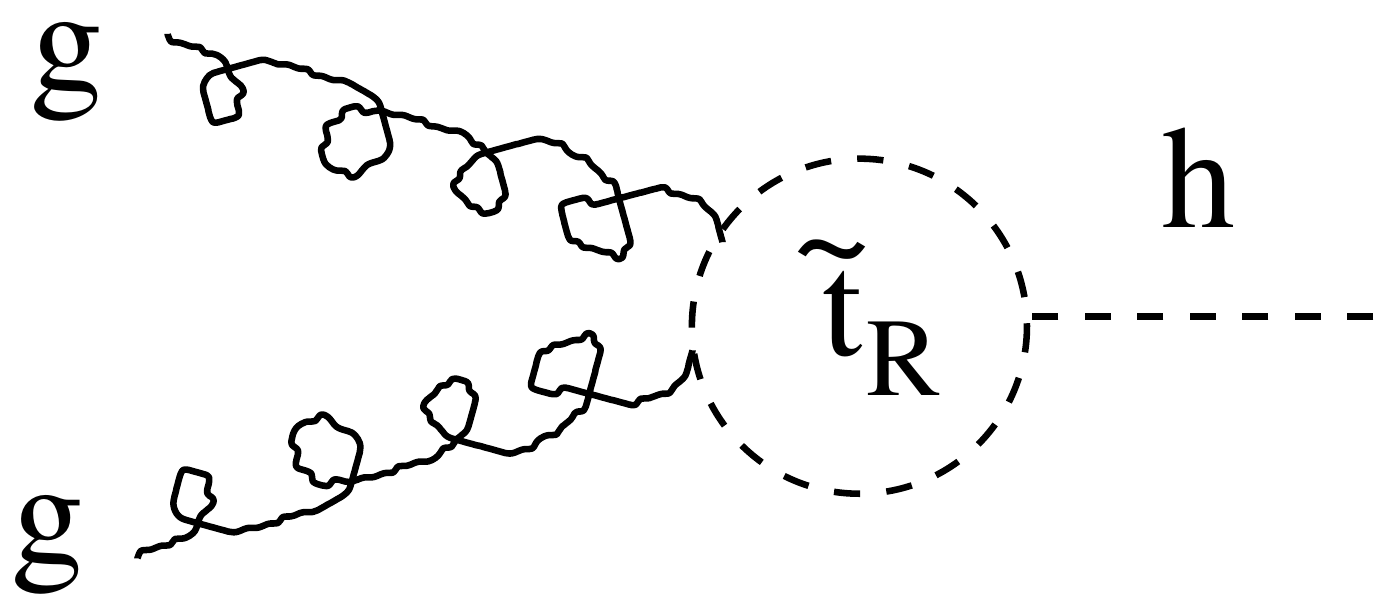}}\hfill{\includegraphics[width=4.5in]{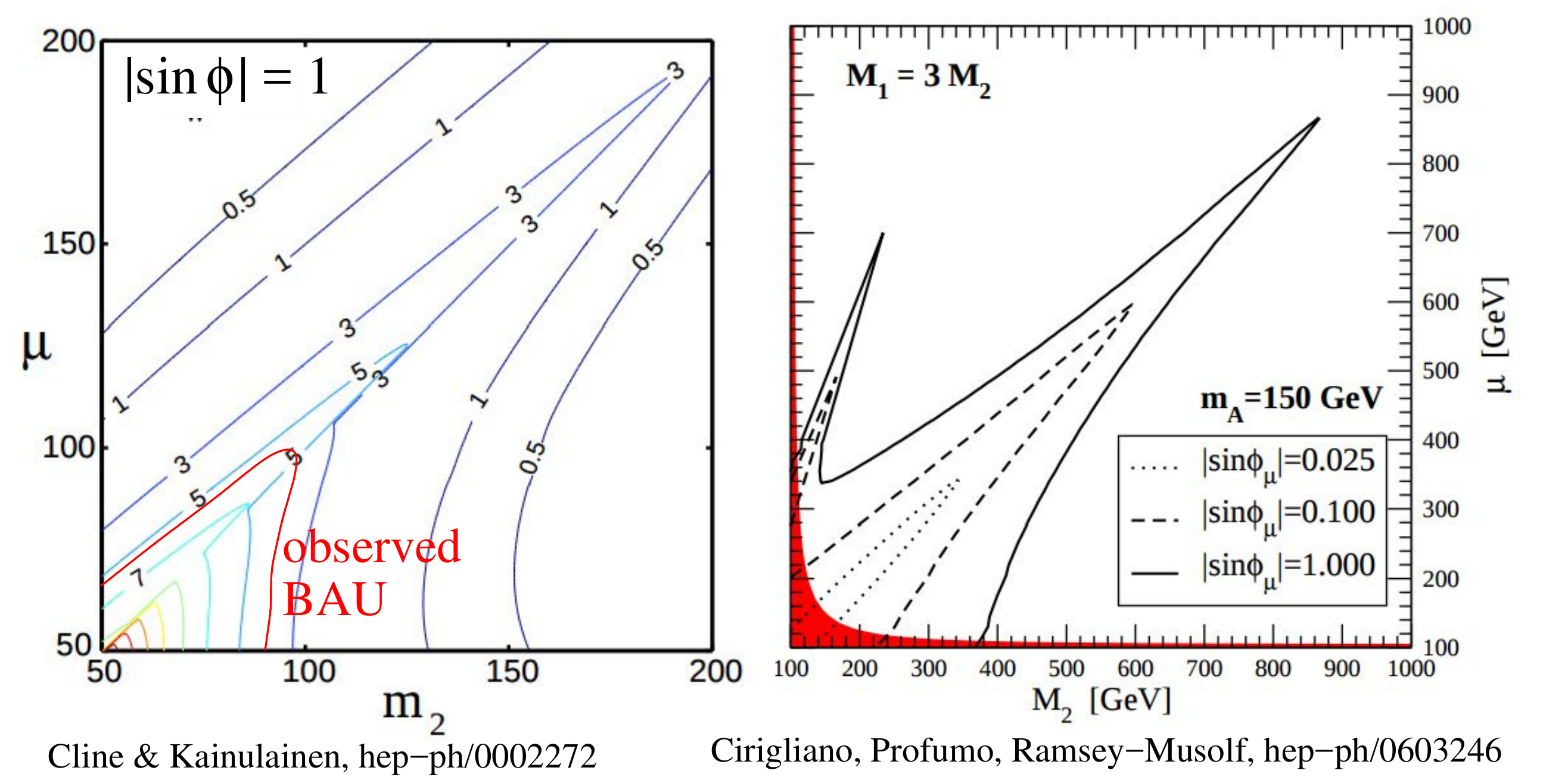}}}
\centerline{(a) $\qquad\qquad\qquad\qquad\qquad\qquad\qquad\qquad\qquad\qquad\qquad$ (b) $\qquad\qquad\qquad\qquad$ }
\caption{(a) MSSM contribution to Higgs production at hadron colliders via gluon fusion with light $\tilde t_\R$ 
(right-handed top squark) in the loop.
(b) Comparison of predictions for the baryon asymmetry versus chargino mass parameters $\mu$ and $m_2$.  Left
plot, using classical chiral force formalism (WKB) \cite{Cline:2000kb}, assumes maximal
CP violating phase $\phi$ and right plot using mass insertion in closed time path formalism, considers varying phase 
\cite{Cirigliano:2006dg}.}
\label{mssm}
\end{figure}
\section{MSSM and two Higgs doublet models}

The Minimal Supersymmetric Standard Model (MSSM) was an initially promising model for EWBG, since having a relatively light
right-handed top squark with mass $m_{\tilde t_\R}\lesssim m_t$ was sufficient for satisfying condition (\ref{vot})  \cite{Carena:1996wj,Cline:1998hy},
and  the CP-violating phase in the chargino mass matrix could lead to a chiral chargino asymmetry, which through interactions
would equilibrate into a chiral quark asymmetry and induce the baryon asymmetry via sphalerons
\cite{Carena:1997gx,Cline:1997vk}. But the difficulties mentioned in section \ref{intro}, in light of increasingly stringent LHC constraints on the
stop mass, as well as EDM constraints, have essentially ruled out this scenario.  The light stop leads to enhanced Higgs 
production via fig.\ \ref{mssm}(a), in conflict with the observed cross section \cite{Cohen:2012zza,Curtin:2012aa}.  
Ref.\ \cite{Carena:2012np} showed that one could
hide the increased cross section if the lightest stable particle had mass $m_\chi< m_h/2$, 
by introducing a large invisible branching ratio for $h\to\chi\chi$ of order $30-60$\%, which however is now ruled out 
\cite{Aad:2015txa,Khachatryan:2016whc}.  Moreover the direct LHC searches for light stops now exclude this scenario
\cite{ATLAS:2017tmd,Sirunyan:2017xse}.
Other finely-tuned loopholes have been pointed out in ref.\ \cite{Liebler:2015ddv}, but have
not generated great enthusiasm in the community, perhaps because of the continuing lack of experimental
evidence for low-energy supersymmetry.

Beyond the difficulty of getting a strong enough phase transition in the MSSM, there is controversy about how to reliably
compute the baryon asymmetry.  Everyone agrees that fluid equations
describing the diffusion of relevant particle species must be solved, to find the spatially dependent chemical potentials of
left-handed fermions with respect to the bubble wall; these determine the rate of biased sphaleron-induced baryon violation. 
The controversy is about how to compute the inhomogeneous source term $S$ that feeds these asymmetries.  It arises from the
CP-violating interactions near the bubble wall.  

Two competing formalisms have emerged for computing $S$.  The
WKB method \cite{Joyce:1994fu,Cline:2000nw} starts with a classical CP-violating force exerted by the wall on particles of 
different chirality, 
\be
	F = \pm {(|m|^2\theta')'\over 2 E^2}
\ee
(see eq.\ (\ref{mths}) below for the definition of $\theta$) and
encodes the chiral charge separation created by this force as the origin of $S$.  The other popular 
method is to make an expansion in powers of the $z$-dependent Higgs VEV in thermal Green's functions, starting
from the closed time path (CTP) formulation of thermal field theory \cite{Riotto:1998zb}, in order to obtain calculable 
expressions.  The WKB formalism, although originally derived from classical dispersion relations,
was also shown to arise starting from CTP \cite{Kainulainen:2001cn,Kainulainen:2002th}, and gives the
leading terms in a systematic expansion in derivatives of the background fields in the bubble wall.\footnote{Source
terms that are one order lower in derivatives can arise when two nearly fermions that mix with each other contribute
to the CP-violating source term in the Boltzmann equations \cite{Konstandin:2005cd}}\ \ 
This approximation is controlled as long as the average de Broglie wavelength $\sim 1/T$ of particles in the plasma 
is small compared to the width $L_w$ of the wall.  In contrast, the expansion in powers of the VEV is not
known to be convergent (though certain subclasses of higher powers can be resummed 
\cite{Carena:2000id,Carena:2002ss}), which may be related to the fact that this formalism 
can predict sizable sources even for masses significantly greater than $T$, despite the expected Boltzmann
suppression.

As a result, the WKB method gives much less optimistic estimates of the baryon asymmetry compared to the VEV
expansion, as fig.\ \ref{mssm}(b) illustrates for the MSSM.  In order to get the observed BAU, ref.\
\cite{Cline:2000kb} needed to assume maximal CP violation ($\phi = {\rm arg}[\mu m_2]=\pi/2)$ in the
chargino mass matrix, as well as light charginos $\mu\sim m_2\sim 100\,$GeV, now ruled out by LHC, whereas
ref.\ \cite{Cirigliano:2006dg} could do so with a phase of order $10^{-2}$ for light charginos, or for
chargino mass $> 800\,$GeV if the phase was maximal.  

EWBG in the next-to-minimal supersymmetric standard model (NMSSM) was shown to have more breathing room in refs.\ 
\cite{Menon:2004wv,Huber:2006wf}, since the extra singlet field could help to strengthen the phase transition as well as
provide new sources of CP violation that are relatively unconstrained by EDMs.  The analysis has been updated in the context
of split SUSY models where the scalar superpartners are much heavier than the neutralinos and charginos, finding positive
results \cite{Demidov:2016wcv}, in models predicting an electron EDM that should be discovered in upcoming searches.
One drawback with the NMSSM is that the extra scalar self-couplings are less protected from running to Landau poles than
those in the MSSM, which are determined by the gauge couplings.

\begin{figure}[t]
\centering
\centerline{\includegraphics[width=3.5in]{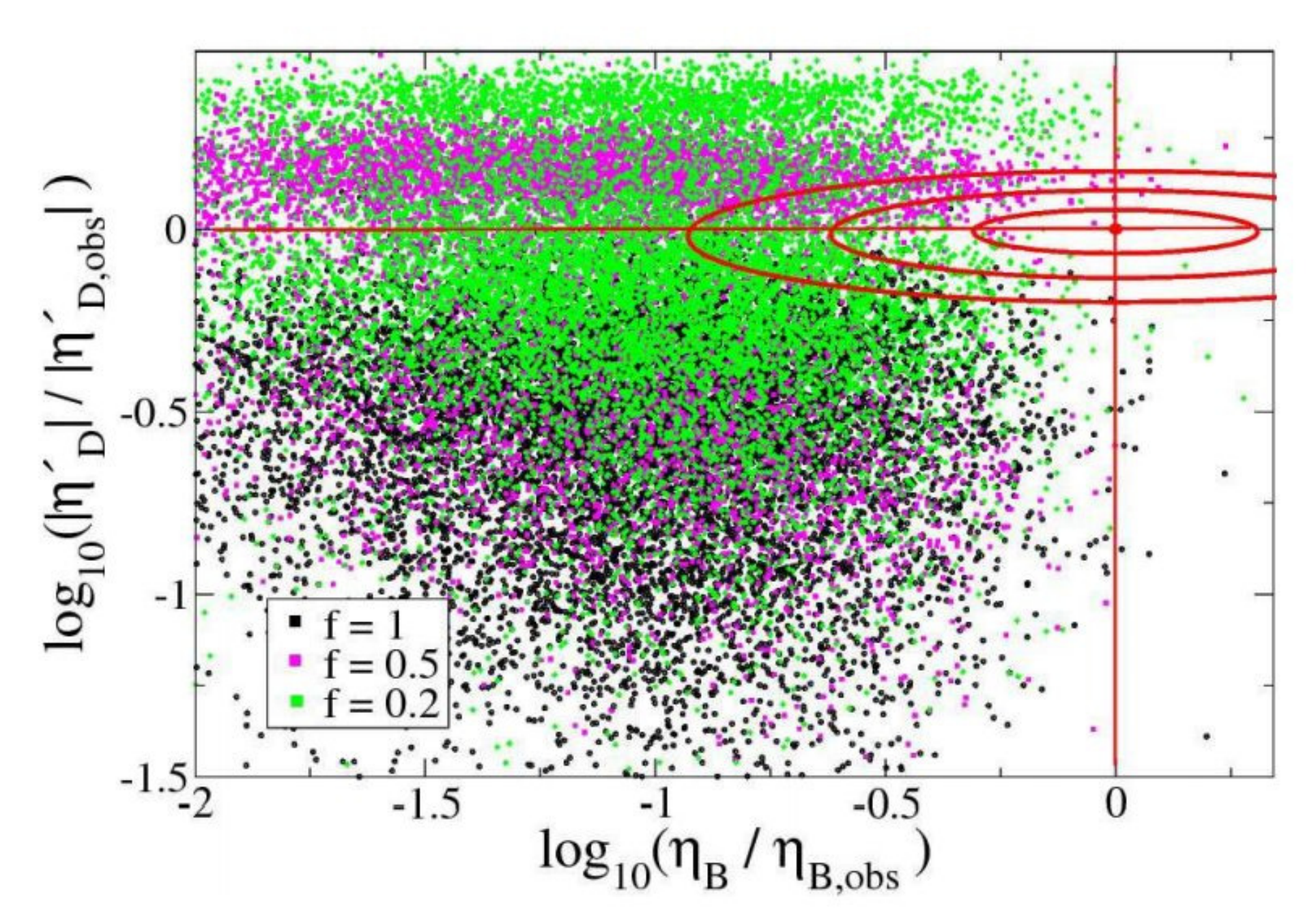}}
\caption{Distribution of two Higgs doublet models passing experimental and consistency constraints, from 
ref.\ \cite{Cline:2011mm}.  Horizontal axis is the predicted BAU in units of the observed value.}
\label{2hdm}
\end{figure}

Electroweak baryogenesis in general two Higgs doublet models (2HDMs) is also in a state of mild controversy.  Ref.\ 
\cite{Cline:2011mm} undertook an extensive study of the allowed parameter space, finding only a small handful of
viable examples in a Monte Carlo Markov chain (MCMC) search yielding 10,000 models.  The result is shown in fig.\ \ref{2hdm}.
In that study, a correlation was sought between the predicted baryon asymmetry (horizontal axis) and possible new sources 
of CP violation in the $b$-quark Yukawa couplings (vertical axis), motivated by the D$\slashed{0}$ like-sign dimuon 
asymmetry, which has since gone away.  As fig.\ \ref{2hdm} shows, no such correlation was found, but for the present argument
all that matters is that very few models exist that predict a large enough BAU.  These few are not very satisfactory, 
because they require such large Higgs self-couplings (to get a strong enough phase transition) that a Landau pole is imminent,
near 1 TeV.  Recently ref.\ \cite{Dorsch:2016nrg} presented a more optimistic outlook for EWBG in 2HDMs.  To understand these
results in light of ref.\ \cite{Cline:2011mm}, it seems likely that the successful models presented there suffer from requiring
very large scalar self-couplings, leading to low-scale Landau poles, as well as very narrow bubble walls, $L_w\sim 1/T$, in
which the derivative expansion assumed for the classical force treatment of the source is not under quantitative control.

\begin{figure}[t]
\centering
\centerline{\includegraphics[width=3.5in]{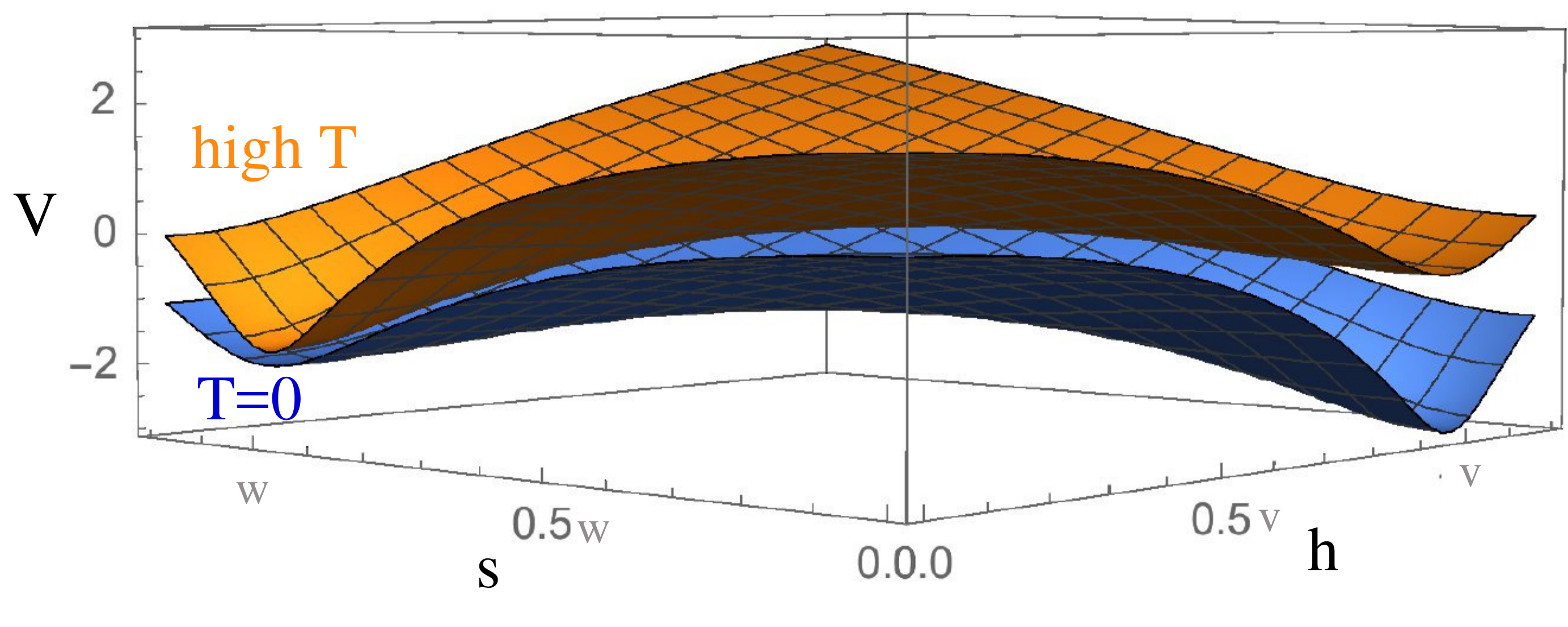}}
\caption{Evolution of sample scalar potential $V(h,s)$ with temperature, illustrating tree-level barrier.}
\label{2fpot}
\end{figure}

\section{Adding a scalar singlet}

It was realized long ago \cite{Choi:1993cv} that coupling the Higgs field to a scalar singlet $S$ can provide a strongly first
order EWPT, if there is already a barrier at tree level (coming from the $(\lambda_m/4)h^2 s^2$ interaction) between the false
vacuum at $(h,s)=(0,w)$ and the true one at $(h,s) = (v,0)$.  There is a two-step transition in the early universe in which the
EWPT is preceded by that where the fully symmetric vacuum $(h,s)=(0,0)$ evolves to a VEV along the $s$ axis.  The second
transition, to the $h$ axis, breaks electroweak symmetry.  The potential for the scalar fields
\be
	V = \sfrac14\lambda_h(h^2-v^2)^2 
+ \sfrac14\lambda_s(s^2-w^2)^2 + \sfrac14\lambda_m h^2 s^2
\ee
 is illustrated in fig.\ 
\ref{2fpot}.  (For simplicity we impose $s\to -s$ symmetry on the potential.)  The transition can easily be very strong since the barrier height is not suppressed by loops or thermal factors.
If the two minima are not too different in height, the small effects of temperature are sufficient to interchange their
relative heights to induce the phase transition.

This idea did not gain popularity immediately since at first it seemed that the cubic terms in the finite-temperature
correction would be sufficient for getting a strong enough transition.  But as the experimental limits that constrain
such contributions have continued to become more stringent, the singlet has become a favored means of boosting the
transition strength, starting with refs.\ \cite{Espinosa:2011ax,Espinosa:2011eu}.  

In ref.\ \cite{Espinosa:2011eu}, it was
realized that the singlet field could also be used to provide a source for the baryon asymmetry, 
by introducing a dimension-5 operator coupling $s$ 
to the usual top quark Yukawa interaction, $(s/\Lambda)\bar Q_3 H(\eta' + i \eta\gamma_5)t_\R$.
  The field-dependent top quark mass then becomes
\be
	m_t(h,s) = { h\over \sqrt{2}}\,\,\left(y_t + (\eta' + i\eta) {s\over\Lambda}\right) \equiv |m_t|e^{i\theta}
\label{mths}
\ee
where $h=h(z)$ and $s=s(z)$ in the bubble wall.  If $\eta$ is nonzero, then there is a CP-violating phase $\theta(z)$.
This is useful for baryogenesis since in the classical force formalism, the source term in the top quark diffusion
equation is proportional to $(|m_t|^2\theta')'$, where $'$ denotes $d/dz$, and $z$ is distance transverse to the bubble wall.

Ref.\ \cite{Cline:2012hg} showed that this also works using the analogous 
dimension-6 operator that is quadratic in $s/\Lambda$, with the advantage
that $s\to -s$ symmetry can be preserved, allowing $s$ to be a dark matter candidate.  It was found to
be easy to generate many models, by a random scan, giving a large enough baryon asymmetry.  To get a strong enough phase
transition, fairly large values of the Higgs-scalar cross coupling $\lambda_{m}\sim 0.5$ are needed, which suppress the 
relic density of $s$ because of the large Higgs-mediated cross section for $ss$ annihilation.  However even though $s$ might
only constitute $\sim 1\%$ of the dark matter, it could still be detected in direct searches due to the correspondingly strong
Higgs-mediated cross section for scattering on nucleons.

A shortcoming of this model, however, is that the scale $\Lambda$ must be rather low, $\lesssim 3\,$TeV, to get
a large enough BAU.  This leads one to question whether the new particles needed to generate the dimension-6 operator
would entail additional constraints from collider searches, and if large couplings leading to low-scale Landau poles might appear in 
a complete model.  One is thus motivated to look for renormalizable models that take advantage of the singlet for enhancing the
phase transition, as well as providing the new source of CP violation needed for EWBG.

\begin{figure}[t]
\centering
\centerline{\includegraphics[width=3.5in]{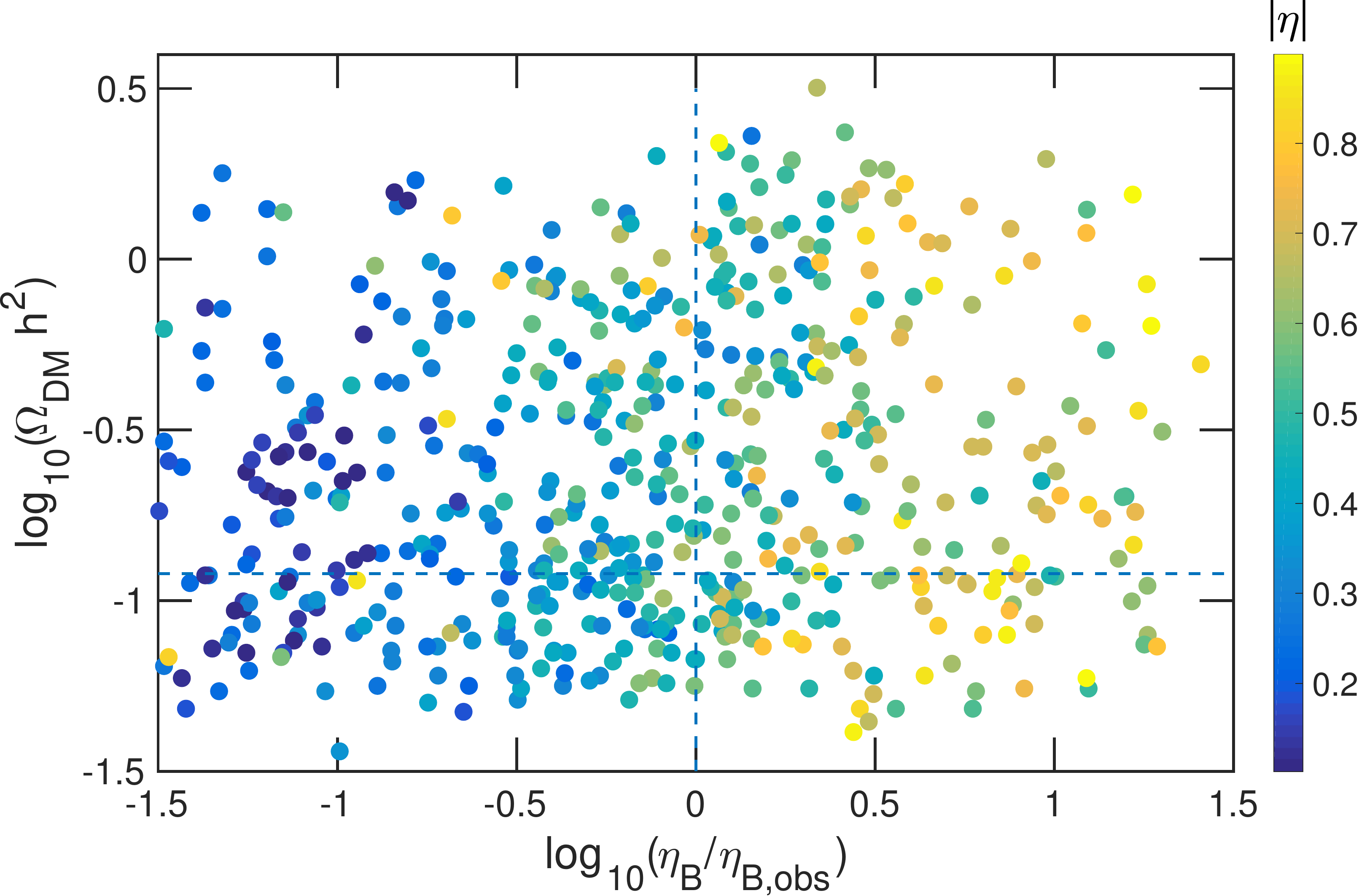}}
\caption{Scatter plot of dark matter density versus baryon asymmetry for the model with dark-matter induced electroweak
baryogenesis.  $\eta$ is the pseudoscalar coupling of the DM to the scalar singlet, eq.\ (\ref{dmlag}).}
\label{scan}
\end{figure}

\section{A more complete model}
\label{complete}
We have proposed a model that overcomes the above-mentioned concerns, by introducing a neutral Majorana fermion $\chi$
that couples to $s$ as \cite{Cline:2017qpe}
\be
	\sfrac12\bar\chi(m_\chi + [\eta' + i\gamma_5\eta] s)\chi
\label{dmlag}
\ee
where $m_\chi, \eta, \eta'$ are real-valued.   This gives a complex mass for $\chi$ in the bubble wall, where
the phase $\theta$ varies with $z$ as in eq.\ (\ref{mths}).  The classical force exerted by the wall thus leads to
the CP asymmetry, in the form of a separation between the two helicity states of $\chi$.  That is not sufficient for
biasing sphalerons since $\chi$ is neutral under SU(2)$_L$.  Thus we require a further interaction for communicating the
CP asymmetry to the standard model doublets.  We introduce an inert Higgs doublet $\phi$ with the {\it CP-portal} interaction
\be
	y_i\bar\chi \phi L_i
\label{portal}
\ee
where $L_i$ is the left-handed doublet of the $i$th generation.  For simplicity we assume that $y_\tau$ is
the dominant coupling, $\eta'=0$, and we neglect possible couplings between $\phi$ and $h$ (especially
\mbox{$(H^\dagger\phi)^2$},
which would induce too-large radiative neutrino masses in conjunction with (\ref{portal})).  Decays and inverse decays
$\phi\leftrightarrow\chi L_\tau$ cause the helicity-asymmetry in $\chi$ to be partially converted to a chemical potential
for $L_\tau$, which then drives the baryon production via sphalerons.

A bonus in this model is that $\chi$ is a good dark matter candidate, which can get the right thermal relic density through
$\chi\bar\chi\to L_\tau\bar L_\tau$ annihilations.  (If $m_\phi< m_\chi$ then $\phi$ would be the dark matter, but since we
assume there is no mass splitting between the neutral components of $\phi$, this would be ruled out by direct detection
constraints on scattering of $\phi$ on nucleons by $Z$ exchange).  The relic density is largely determined by $y_\tau$, which
also has a strong impact on the BAU, making the model more constrained.  Nevertheless, we find many models with reasonable
values of the parameters ($\lambda_m\sim\eta\sim 0.5$, $y_\tau\sim 0.6$, $m_\chi \sim 50\,$GeV, $m_\phi\sim 120\,$GeV,
$m_s\sim 110\,$GeV) 
that are consistent with the observed BAU and dark matter density.  The results of a random scan
are shown in fig.\ \ref{scan}.  In contrast to the analogous result fig.\ \ref{2hdm} for 2HDMs, where MCMC was needed to find
the few viable models, here no great effort is required to generate successful examples.  

\begin{figure}[t]
\centering
\centerline{\includegraphics[width=4.5in]{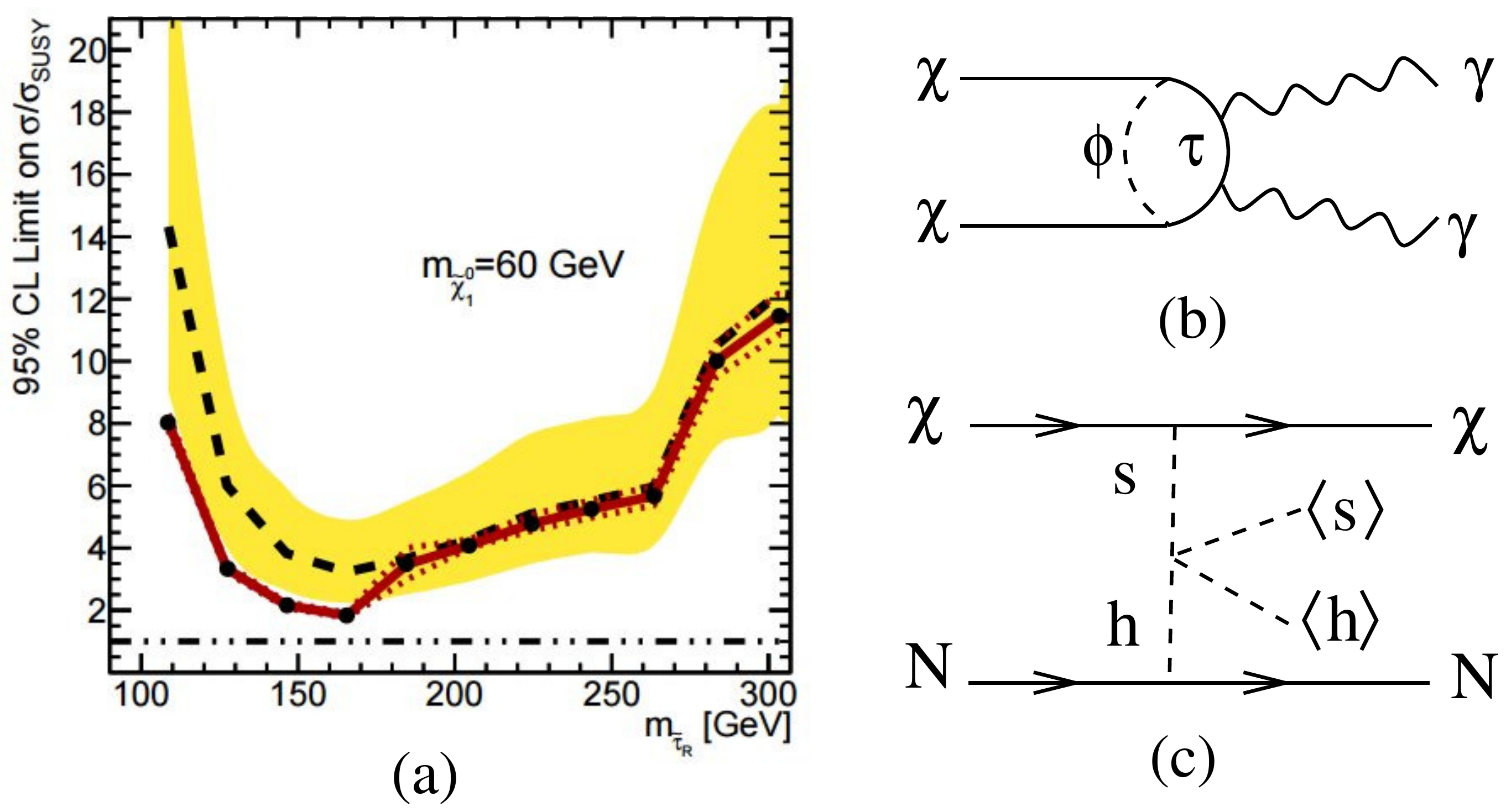}}
\caption{(a) ATLAS upper limit on $\tilde\tau$ pair production cross section in the MSSM for lightest superpartner (dark matter)
mass of 60\,GeV from ref.\ \cite{Aad:2014yka}.  Dash-dotted line is the predicted cross section.  We can reinterpret $\tilde\tau$ as 
$\phi$
and the neutralino as $\chi$ to apply the constraints to our model.  (b) Diagram for producing gamma ray lines from dark matter
annihilation.  (c) Diagram generating DM scattering on nuclei, in the presence of nonvanishing $\langle s\rangle$ VEV.}
\label{detect}
\end{figure}

This model has strong potential for discovery at LHC.  The Drell-Yan production of $\phi^\pm$ pairs, followed by $\phi\to\tau\chi$
decays, is similar to $\tilde\tau$ pair production followed by $\tilde\tau\to\tau\chi_0$ in the MSSM.  Fig.\ \ref{detect}
shows the ATLAS limit from Run 1 on the production cross section, versus the predicted cross section, as a function of 
$m_{\tilde\tau_\R}$ and neutralino mass of 60\,GeV.  To within factors of 2 (since ATLAS considers combined
production of $\tilde\tau_\R$ and $\tilde\tau_\L$ pairs, whereas $\phi$ has no right-handed counterpart), 
these limits also apply to our model.  They indicate that for $m_\phi\sim 120\,$GeV and $m_\chi\sim 60\,$GeV 
(as predicted by our model), the limiting cross section is only a few times greater than the predicted one, giving hope that
detection could be possible with the Run 2 data.

There is also potential for indirect detection through the emission of gamma ray lines from  $\chi\bar\chi \to
\gamma\gamma$, from the diagram of fig.\ \ref{detect}(b).  Unlike the tree-level annihilation $\chi\bar\chi\to \tau\bar\tau$
which is $p$-waved suppressed, this process is $s$-wave, with $\langle\sigma v\rangle\cong 4\times 10^{-30}$cm$^3/$s.  This
is not far below the most optimistic constraint $\sim 10^{-30}$cm$^3/$s (depending upon assumptions about the DM density profile in the galactic
center) from Fermi/LAT \cite{Ackermann:2015lka}.  

For direct detection, the cross section is unobservably small unless we allow
for a small VEV $\langle s\rangle$ at zero temperature.  Then singlet-Higgs mixing gives rise to the diagram of fig.\
\ref{detect}(c).  Current bounds from direct searches limit the mixing angle at the level $\theta_{hs} < 0.04$.  In fact, some small
amount of mixing is required for successful baryogenesis in this model, since if $s\to-s$ is an exact symmetry of the scalar
potential, then the early universe will be equally populated by domains with $s>0$ and $s<0$ during the EWPT, containing
equal and opposite values of the BAU that will eventually average to zero.  Only very small (Planck-suppressed) mixing is
needed to avoid this problem: lifting the degeneracy between the $s<0$ and $s>0$ false vacua before the EWPT will eliminate
the higher energy phase as long as the domain walls separating the two phases annihilate faster than the Hubble rate.
Hence we do not expect that relaxing this simplifying approximation will change our quantitative
estimates of the BAU.  

\section{Outlook}

The model described in section \ref{complete} is one example showing that electroweak baryogenesis,
far from being dead, can be achieved in renormalizable models that are within the reach of discovery
at LHC, without requiring fine tuning or unreasonably large dimensionless couplings.  It is likely that
many such models exist, that take advantage of a scalar singlet to get a strong EWPT and relatively 
unconstrained CP violation.

For example, one can easily generate eq.\ (\ref{mths}) by integrating out a heavy 
vectorlike isosinglet top partner $T$, with interactions
\be
	\eta\, \bar t_R S T_L + M \bar T_L T_R + y' \bar T_R H t_L
\ee
This gives a nonstandard contribution to the top quark mass in the bubble wall, $(\eta y'/M)\, \bar t_R S H t_L$ that
operates just like eq.\ (\ref{mths}) to produce a CP asymmetry, if $\eta y'$ has a phase relative to the SM top Yukawa
coupling.  The current mass limit on vector-like top partners is around $1\,$TeV \cite{ATLAS:2016qlg}, making this an
excellent candidate for testable new physics that can give the baryon asymmetry.  Work on this is in progress.

\begin{figure}[t]
\centering
\centerline{\includegraphics[width=2in]{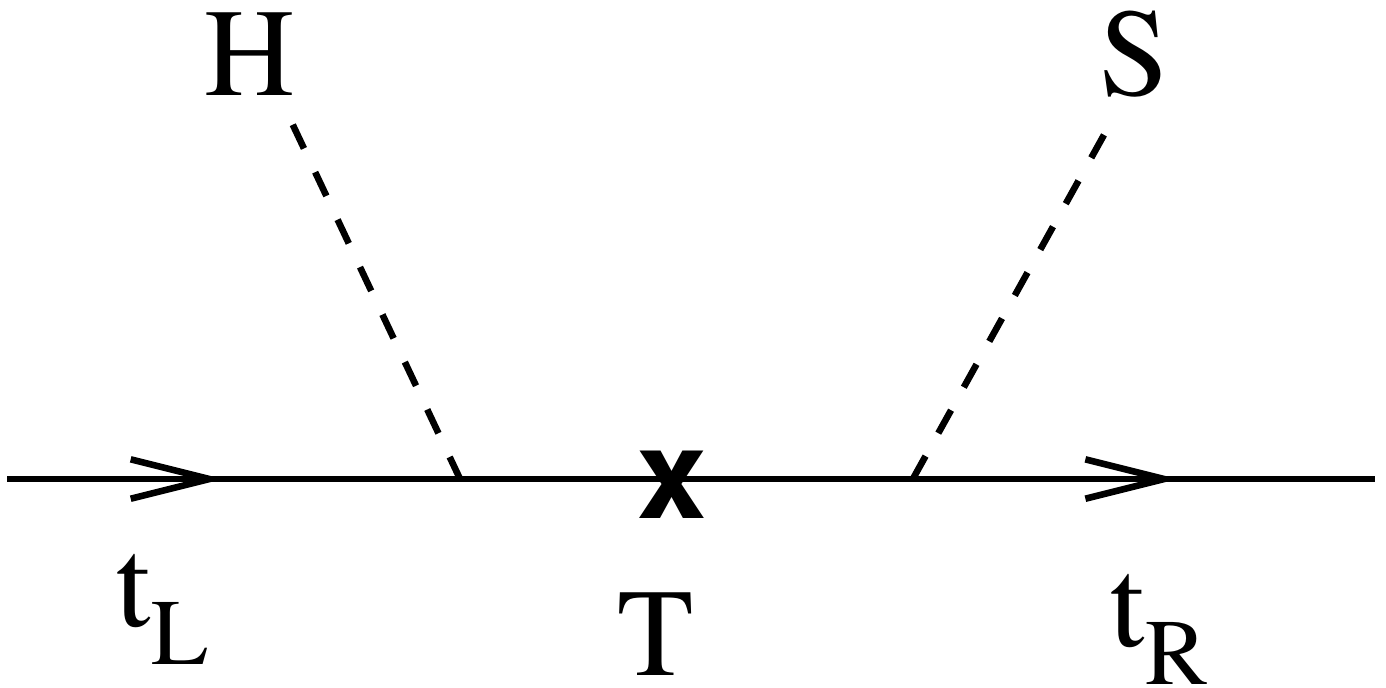}}
\caption{Generation of dimension-5 contribution to top quark mass, $\bar Q_3 H S t_\R/\Lambda$, through a heavy vectorlike top
partner.}
\label{top-partner}
\end{figure}

\enlargethispage{20pt}

\funding{This work was supported by the Natural Sciences and Engineering Research Council (Canada)
and Fonds de recherche du Qu\'bec---Nature et technologies.}

\ack{I thank Kimmo Kainulainen and David Tucker-Smith for their collaboration on this work.}



\begin{thebibliography}{9}

\bibitem{Peon:1988kx} 
  B.~Peon,
  ``Is Hinchliffe's Rule True?,''
  Submitted to: Annals of Gnosis (1988)

\bibitem{Pilaftsis:2003gt} 
  A.~Pilaftsis and T.~E.~J.~Underwood,
  ``Resonant leptogenesis,''
  Nucl.\ Phys.\ B {\bf 692}, 303 (2004)
  doi:10.1016/j.nuclphysb.2004.05.029
  [hep-ph/0309342].

\bibitem{Baldes:2016gaf} 
  I.~Baldes, T.~Konstandin and G.~Servant,
  ``Flavor Cosmology: Dynamical Yukawas in the Froggatt-Nielsen Mechanism,''
  JHEP {\bf 1612}, 073 (2016)
  doi:10.1007/JHEP12(2016)073
  [arXiv:1608.03254 [hep-ph]].

\bibitem{Patel:2012pi} 
  H.~H.~Patel and M.~J.~Ramsey-Musolf,
  ``Stepping Into Electroweak Symmetry Breaking: Phase Transitions and Higgs Phenomenology,''
  Phys.\ Rev.\ D {\bf 88}, 035013 (2013)
  doi:10.1103/PhysRevD.88.035013
  [arXiv:1212.5652 [hep-ph]].

\bibitem{Blinov:2015sna} 
  N.~Blinov, J.~Kozaczuk, D.~E.~Morrissey and C.~Tamarit,
  ``Electroweak Baryogenesis from Exotic Electroweak Symmetry Breaking,''
  Phys.\ Rev.\ D {\bf 92}, no. 3, 035012 (2015)
  doi:10.1103/PhysRevD.92.035012
  [arXiv:1504.05195 [hep-ph]].

\bibitem{Inoue:2015pza} 
  S.~Inoue, G.~Ovanesyan and M.~J.~Ramsey-Musolf,
  ``Two-Step Electroweak Baryogenesis,''
  Phys.\ Rev.\ D {\bf 93}, 015013 (2016)
  doi:10.1103/PhysRevD.93.015013
  [arXiv:1508.05404 [hep-ph]].

\bibitem{Kuzmin:1985mm} 
  V.~A.~Kuzmin, V.~A.~Rubakov and M.~E.~Shaposhnikov,
  ``On the Anomalous Electroweak Baryon Number Nonconservation in the Early Universe,''
  Phys.\ Lett.\  {\bf 155B}, 36 (1985).
  doi:10.1016/0370-2693(85)91028-7

\bibitem{Cohen:1990it} 
  A.~G.~Cohen, D.~B.~Kaplan and A.~E.~Nelson,
  ``Baryogenesis at the weak phase transition,''
  Nucl.\ Phys.\ B {\bf 349}, 727 (1991).
  doi:10.1016/0550-3213(91)90395-E

\bibitem{Turok:1990in} 
  N.~Turok and J.~Zadrozny,
  ``Dynamical generation of baryons at the electroweak transition,''
  Phys.\ Rev.\ Lett.\  {\bf 65}, 2331 (1990).
  doi:10.1103/PhysRevLett.65.2331

\bibitem{Moore:1998swa} 
  G.~D.~Moore,
  ``Measuring the broken phase sphaleron rate nonperturbatively,''
  Phys.\ Rev.\ D {\bf 59}, 014503 (1999)
  doi:10.1103/PhysRevD.59.014503
  [hep-ph/9805264].

\bibitem{Moore:1995ua} 
  G.~D.~Moore and T.~Prokopec,
  ``Bubble wall velocity in a first order electroweak phase transition,''
  Phys.\ Rev.\ Lett.\  {\bf 75}, 777 (1995)
  doi:10.1103/PhysRevLett.75.777
  [hep-ph/9503296].

\bibitem{Kozaczuk:2015owa} 
  J.~Kozaczuk,
  ``Bubble Expansion and the Viability of Singlet-Driven Electroweak Baryogenesis,''
  JHEP {\bf 1510}, 135 (2015)
  doi:10.1007/JHEP10(2015)135
  [arXiv:1506.04741 [hep-ph]].

\bibitem{Kurup:2017dzf} 
  G.~Kurup and M.~Perelstein,
  ``Dynamics of Electroweak Phase Transition In Singlet-Scalar Extension of the Standard Model,''
  arXiv:1704.03381 [hep-ph].

\bibitem{Bodeker:2009qy} 
  D.~Bodeker and G.~D.~Moore,
  ``Can electroweak bubble walls run away?,''
  JCAP {\bf 0905}, 009 (2009)
  doi:10.1088/1475-7516/2009/05/009
  [arXiv:0903.4099 [hep-ph]].

\bibitem{Bodeker:2017cim} 
  D.~Bodeker and G.~D.~Moore,
  ``Electroweak Bubble Wall Speed Limit,''
  JCAP {\bf 1705}, no. 05, 025 (2017)
  doi:10.1088/1475-7516/2017/05/025
  [arXiv:1703.08215 [hep-ph]].



\bibitem{Carena:1996wj} 
  M.~Carena, M.~Quiros and C.~E.~M.~Wagner,
  ``Opening the window for electroweak baryogenesis,''
  Phys.\ Lett.\ B {\bf 380}, 81 (1996)
  doi:10.1016/0370-2693(96)00475-3
  [hep-ph/9603420].

\bibitem{Cline:1998hy} 
  J.~M.~Cline and G.~D.~Moore,
  ``Supersymmetric electroweak phase transition: Baryogenesis versus experimental constraints,''
  Phys.\ Rev.\ Lett.\  {\bf 81}, 3315 (1998)
  doi:10.1103/PhysRevLett.81.3315
  [hep-ph/9806354].

\bibitem{Carena:1997gx} 
  M.~Carena, M.~Quiros, A.~Riotto, I.~Vilja and C.~E.~M.~Wagner,
  ``Electroweak baryogenesis and low-energy supersymmetry,''
  Nucl.\ Phys.\ B {\bf 503}, 387 (1997)
  doi:10.1016/S0550-3213(97)00412-4
  [hep-ph/9702409].

\bibitem{Cline:1997vk} 
  J.~M.~Cline, M.~Joyce and K.~Kainulainen,
  ``Supersymmetric electroweak baryogenesis in the WKB approximation,''
  Phys.\ Lett.\ B {\bf 417}, 79 (1998)
  Erratum: [Phys.\ Lett.\ B {\bf 448}, 321 (1999)]
  doi:10.1016/S0370-2693(99)00033-7, 10.1016/S0370-2693(97)01361-0
  [hep-ph/9708393].

\bibitem{Cohen:2012zza} 
  T.~Cohen, D.~E.~Morrissey and A.~Pierce,
  ``Electroweak Baryogenesis and Higgs Signatures,''
  Phys.\ Rev.\ D {\bf 86}, 013009 (2012)
  doi:10.1103/PhysRevD.86.013009
  [arXiv:1203.2924 [hep-ph]].

\bibitem{Curtin:2012aa} 
  D.~Curtin, P.~Jaiswal and P.~Meade,
  ``Excluding Electroweak Baryogenesis in the MSSM,''
  JHEP {\bf 1208}, 005 (2012)
  doi:10.1007/JHEP08(2012)005
  [arXiv:1203.2932 [hep-ph]].

\bibitem{Carena:2012np} 
  M.~Carena, G.~Nardini, M.~Quiros and C.~E.~M.~Wagner,
  ``MSSM Electroweak Baryogenesis and LHC Data,''
  JHEP {\bf 1302}, 001 (2013)
  doi:10.1007/JHEP02(2013)001
  [arXiv:1207.6330 [hep-ph]].

\bibitem{Aad:2015txa} 
  G.~Aad {\it et al.} [ATLAS Collaboration],
  ``Search for invisible decays of a Higgs boson using vector-boson fusion in $pp$ collisions at $\sqrt{s}=8$ TeV with the ATLAS detector,''
  JHEP {\bf 1601}, 172 (2016)
  doi:10.1007/JHEP01(2016)172
  [arXiv:1508.07869 [hep-ex]].


\bibitem{Khachatryan:2016whc} 
  V.~Khachatryan {\it et al.} [CMS Collaboration],
  ``Searches for invisible decays of the Higgs boson in pp collisions at sqrt(s) = 7, 8, and 13 TeV,''
  JHEP {\bf 1702}, 135 (2017)
  doi:10.1007/JHEP02(2017)135
  [arXiv:1610.09218 [hep-ex]].

\bibitem{ATLAS:2017tmd} 
  The ATLAS collaboration [ATLAS Collaboration],
  ``Search for direct top squark pair production in final states with two leptons in $\sqrt{s} = 13$ TeV $pp$ collisions with the ATLAS detector,''
  ATLAS-CONF-2017-034.

\bibitem{Sirunyan:2017xse} 
  A.~M.~Sirunyan {\it et al.} [CMS Collaboration],
  ``Search for top squark pair production in pp collisions at sqrt(s)=13 TeV using single lepton events,''
  arXiv:1706.04402 [hep-ex].

\bibitem{Liebler:2015ddv} 
  S.~Liebler, S.~Profumo and T.~Stefaniak,
  ``Light Stop Mass Limits from Higgs Rate Measurements in the MSSM: Is MSSM Electroweak Baryogenesis Still Alive After All?,''
  JHEP {\bf 1604}, 143 (2016)
  doi:10.1007/JHEP04(2016)143
  [arXiv:1512.09172 [hep-ph]].

\bibitem{Joyce:1994fu} 
  M.~Joyce, T.~Prokopec and N.~Turok,
  ``Electroweak baryogenesis from a classical force,''
  Phys.\ Rev.\ Lett.\  {\bf 75}, 1695 (1995)
  Erratum: [Phys.\ Rev.\ Lett.\  {\bf 75}, 3375 (1995)]
  doi:10.1103/PhysRevLett.75.1695
  [hep-ph/9408339].

\bibitem{Cline:2000nw} 
  J.~M.~Cline, M.~Joyce and K.~Kainulainen,
  ``Supersymmetric electroweak baryogenesis,''
  JHEP {\bf 0007}, 018 (2000)
  doi:10.1088/1126-6708/2000/07/018
  [hep-ph/0006119].

\bibitem{Riotto:1998zb} 
  A.~Riotto,
  ``The More relaxed supersymmetric electroweak baryogenesis,''
  Phys.\ Rev.\ D {\bf 58}, 095009 (1998)
  doi:10.1103/PhysRevD.58.095009
  [hep-ph/9803357].

\bibitem{Kainulainen:2001cn} 
  K.~Kainulainen, T.~Prokopec, M.~G.~Schmidt and S.~Weinstock,
  ``First principle derivation of semiclassical force for electroweak baryogenesis,''
  JHEP {\bf 0106}, 031 (2001)
  doi:10.1088/1126-6708/2001/06/031
  [hep-ph/0105295].

\bibitem{Kainulainen:2002th} 
  K.~Kainulainen, T.~Prokopec, M.~G.~Schmidt and S.~Weinstock,
  ``Semiclassical force for electroweak baryogenesis: Three-dimensional derivation,''
  Phys.\ Rev.\ D {\bf 66}, 043502 (2002)
  doi:10.1103/PhysRevD.66.043502
  [hep-ph/0202177].

\bibitem{Konstandin:2005cd} 
  T.~Konstandin, T.~Prokopec, M.~G.~Schmidt and M.~Seco,
  ``MSSM electroweak baryogenesis and flavor mixing in transport equations,''
  Nucl.\ Phys.\ B {\bf 738}, 1 (2006)
  doi:10.1016/j.nuclphysb.2005.11.028
  [hep-ph/0505103].

\bibitem{Carena:2000id} 
  M.~Carena, J.~M.~Moreno, M.~Quiros, M.~Seco and C.~E.~M.~Wagner,
  ``Supersymmetric CP violating currents and electroweak baryogenesis,''
  Nucl.\ Phys.\ B {\bf 599}, 158 (2001)
  doi:10.1016/S0550-3213(01)00032-3
  [hep-ph/0011055].

\bibitem{Carena:2002ss} 
  M.~Carena, M.~Quiros, M.~Seco and C.~E.~M.~Wagner,
  ``Improved results in supersymmetric electroweak baryogenesis,''
  Nucl.\ Phys.\ B {\bf 650}, 24 (2003)
  doi:10.1016/S0550-3213(02)01065-9
  [hep-ph/0208043].


\bibitem{Cline:2000kb} 
  J.~M.~Cline and K.~Kainulainen,
  ``A New source for electroweak baryogenesis in the MSSM,''
  Phys.\ Rev.\ Lett.\  {\bf 85}, 5519 (2000)
  doi:10.1103/PhysRevLett.85.5519
  [hep-ph/0002272].


\bibitem{Cirigliano:2006dg} 
  V.~Cirigliano, S.~Profumo and M.~J.~Ramsey-Musolf,
  ``Baryogenesis, Electric Dipole Moments and Dark Matter in the MSSM,''
  JHEP {\bf 0607}, 002 (2006)
  doi:10.1088/1126-6708/2006/07/002
  [hep-ph/0603246].


\bibitem{Menon:2004wv} 
  A.~Menon, D.~E.~Morrissey and C.~E.~M.~Wagner,
  ``Electroweak baryogenesis and dark matter in the nMSSM,''
  Phys.\ Rev.\ D {\bf 70}, 035005 (2004)
  doi:10.1103/PhysRevD.70.035005
  [hep-ph/0404184].

\bibitem{Huber:2006wf} 
  S.~J.~Huber, T.~Konstandin, T.~Prokopec and M.~G.~Schmidt,
  ``Electroweak Phase Transition and Baryogenesis in the nMSSM,''
  Nucl.\ Phys.\ B {\bf 757}, 172 (2006)
  doi:10.1016/j.nuclphysb.2006.09.003
  [hep-ph/0606298].

\bibitem{Demidov:2016wcv} 
  S.~V.~Demidov, D.~S.~Gorbunov and D.~V.~Kirpichnikov,
  ``Split NMSSM with electroweak baryogenesis,''
  JHEP {\bf 1611}, 148 (2016)
  doi:10.1007/JHEP11(2016)148
  [arXiv:1608.01985 [hep-ph]].

\bibitem{Cline:2011mm} 
  J.~M.~Cline, K.~Kainulainen and M.~Trott,
  ``Electroweak Baryogenesis in Two Higgs Doublet Models and B meson anomalies,''
  JHEP {\bf 1111}, 089 (2011)
  doi:10.1007/JHEP11(2011)089
  [arXiv:1107.3559 [hep-ph]].

\bibitem{Dorsch:2016nrg} 
  G.~C.~Dorsch, S.~J.~Huber, T.~Konstandin and J.~M.~No,
  ``A Second Higgs Doublet in the Early Universe: Baryogenesis and Gravitational Waves,''
  arXiv:1611.05874 [hep-ph].

\bibitem{Choi:1993cv} 
  J.~Choi and R.~R.~Volkas,
  ``Real Higgs singlet and the electroweak phase transition in the Standard Model,''
  Phys.\ Lett.\ B {\bf 317}, 385 (1993)
  doi:10.1016/0370-2693(93)91013-D
  [hep-ph/9308234].

\bibitem{Espinosa:2011ax} 
  J.~R.~Espinosa, T.~Konstandin and F.~Riva,
  ``Strong Electroweak Phase Transitions in the Standard Model with a Singlet,''
  Nucl.\ Phys.\ B {\bf 854}, 592 (2012)
  [arXiv:1107.5441 [hep-ph]].

\bibitem{Espinosa:2011eu} 
  J.~R.~Espinosa, B.~Gripaios, T.~Konstandin and F.~Riva,
  ``Electroweak Baryogenesis in Non-minimal Composite Higgs Models,''
  JCAP {\bf 1201}, 012 (2012)
  [arXiv:1110.2876 [hep-ph]].

\bibitem{Cline:2012hg} 
  J.~M.~Cline and K.~Kainulainen,
  ``Electroweak baryogenesis and dark matter from a singlet Higgs,''
  JCAP {\bf 1301}, 012 (2013)
  doi:10.1088/1475-7516/2013/01/012
  [arXiv:1210.4196 [hep-ph]].

\bibitem{Cline:2017qpe} 
  J.~M.~Cline, K.~Kainulainen and D.~Tucker-Smith,
  ``Electroweak baryogenesis from a dark sector,''
  arXiv:1702.08909 [hep-ph].

\bibitem{Aad:2014yka} 
  G.~Aad {\it et al.} [ATLAS Collaboration],
  ``Search for the direct production of charginos, neutralinos and staus in final states with at least two hadronically decaying taus and missing transverse momentum in $pp$ collisions at $\sqrt{s}$ = 8 TeV with the ATLAS detector,''
  JHEP {\bf 1410}, 096 (2014)
  doi:10.1007/JHEP10(2014)096
  [arXiv:1407.0350 [hep-ex]].

\bibitem{Ackermann:2015lka} 
  M.~Ackermann {\it et al.} [Fermi-LAT Collaboration],
  ``Updated search for spectral lines from Galactic dark matter interactions with pass 8 data from the Fermi Large Area Telescope,''
  Phys.\ Rev.\ D {\bf 91}, no. 12, 122002 (2015)
  doi:10.1103/PhysRevD.91.122002
  [arXiv:1506.00013 [astro-ph.HE]].

\bibitem{ATLAS:2016qlg} 
  The ATLAS collaboration [ATLAS Collaboration],
  ``Search for pair production of vector-like top partners in events with exactly one lepton and large missing transverse momentum in $\sqrt{s}=13$ TeV $pp$ collisions with the ATLAS detector,''
  ATLAS-CONF-2016-101.

\end{thebibliography}
\end{document}